# Super-resolution and signal separation in contact Kelvin probe force microscopy of electrochemically active ferroelectric materials

Maxim Ziatdinov,[1,3] Dohyung Kim,[2] Sabine Neumayer,[1] Liam Collins,[1] Mahshid Ahmadi,[2] Rama K. Vasudevan,[1] Stephen Jesse[1], Myung Hyun Ann[4], Jong H. Kim[4], and Sergei V. Kalinin[1]

[1] The Center for Nanophase Materials Sciences, Oak Ridge National Laboratory, Oak Ridge, TN 37831, USA

[2] Joint Institute for Advanced Materials, Department of Materials Science and Engineering, University of Tennessee, Knoxville, TN 37996, USA

[3] Computational Sciences and Engineering Division, Oak Ridge National Laboratory, Oak Ridge, TN 37831, USA

[4] Department of Molecular Science and Technology, Ajou University, Suwon 16499, Republic of Korea.

**Abstract**

Imaging mechanisms in contact Kelvin Probe Force Microscopy (cKPFM) are explored via information theory-based methods. Gaussian Processes are used to achieve super-resolution in the cKPFM signal, effectively extrapolating across the spatial and parameter space. Tensor factorization is applied to reduce the multidimensional signal to the tensor convolution of the scalar functions that show clear trending behavior with the imaging parameters. These methods establish a workflow for the analysis of the multidimensional data sets, that can then be related to the relevant physical mechanisms. We also provide an interactive Google Colab notebook that goes through all the analysis discussed in the paper.

**Introduction**

Understanding of ferroelectric and electrochemical materials alike necessitates untangling coupled dynamics of polarization and electrochemical dipoles as well as mobile ionic and electronic charges.[1-7] In ferroelectrics, this includes the dynamics of unit-cell level polarization dipoles defined on the unit cell levels, as well as physics and chemistry of polarization screening phenomena at surfaces and interfaces.[5] In electrochemically active materials, this includes the electrochemical polarization, surface and bulk electrochemical reactions and interfacial ion transfer.[8-10] Applications such as ferroelectric non-volatile memories, storage, and tunneling devices, batteries and fuel cells, electrochemical sensors and memristors all necessitate these mechanisms be probed on the nanometer scale level of individual grains, structural and topological defects.[11]

While traditionally difficult to study via classical electrochemical, scattering, and optical probes, these phenomena are invariably linked to electromechanical responses. Correspondingly, the measurements based on detection of local electromechanical coupling provide the pathway for exploring ferroelectric and electrochemical phenomena on the nanometer scales, via Piezoresponse Force Microscopy (PFM)[12-14] and Electrochemical Strain Microscopy (ESM)[15-17] and associated spectroscopies. In PFM and ESM, the application of the bias to the local probe in contact with the surface results in strong, ~GV/m, highly localized electric fields. The associated surface deformations and contact electrostatic forces result in probe displacements that can be detected on the ~pm level, with the spatial resolution determined by the localization of the electric field at the tip-surface junction to ~10 nm level.

Originally, PFM was developed as a technique for probing domain structure and polarization dynamics in strongly ferroelectric materials with the electromechanical responses in the 10-100 pm/V range.[18-20] In these materials, the electromechanical coupling is primarily due to the inverse piezoelectric effect, allowing for unambiguous interpretation of the imaging and spectroscopic data. At the same time, advances in sensitivity of PFM and ESM instrumentation and emergence of resonance-enhanced modes such as dual amplitude resonance tracking (DART)[21] and band excitation (BE)[22] have allowed exploring materials with low (1-10 pm/V) electromechanical responses, including ionic conductors and electrochemically active systems. In this case, the increased sensitivity allowed detection of multiple responses contributing to measured signals. For example, in ferroelectrics both the polarization and surface electrochemistry contribute to measured electromechanical signal.[1] Similarly, responses in ionic conductors comprise the contribution from surface and bulk ionic motion via electrostrictive coupling, electrostatic forces, and Vegard strains.[23, 24] This multitude of contributing mechanisms necessitates the development of experimental strategies for separating possible electromechanical coupling, and analysis of relevant materials phenomena.

To address this challenge, a number of spectroscopic techniques probing the response dynamic in time- and voltage domains were proposed. These include 4D spectroscopic modes such as band excitation switching spectroscopy PFM and amplitude measurements,[25] as well as more complex 5D modes including first order reversal curve measurements,[26] dynamic PFM,[27] and contact Kelvin Probe Force Microscopy (cKPFM).[28] In these spectroscopic imaging techniques, the time- and voltage dependence of electromechanical signal is used to separate fast local polarization dynamics from slower ionic processes and mesoscopic polarization dynamics.

However, implementation of high-dimensional spectroscopic modes brings obvious challenges. One is that of measurement time, where the data acquisition rate is limited by the cantilever/detector bandwidth and imaging time is limited both by the drift stability of microscope system and the fact that a probe may not survive multiple hours of high-voltage spectroscopy While 4D PFM and ESM spectroscopic imaging modes generally allow imaging with high sampling of spatial and spectroscopic domain, the 5D modes are severely limited. Here, sampling the 3D parameter space necessitates ~1-10 s per pixel, which for the spatial 60x60 pixels grids leads to several hours acquisition. Correspondingly, of interest is the development of algorithms that allow to efficiently augment the data in high dimensional parameter space, efficiently interpolating and potentially enabling super-resolution imaging in spatial and parameter space. Second, the high dimensional spectroscopic data sets necessitate the development of the tools for dimensionality reduction, compressing the high dimensional data sets to low-dimensional representations that allow data compression, exploratory data analysis, and potentially enable separation of relevant mechanisms.

Here, we demonstrate the use of the Gaussian Process (GP) regression for achieving the super resolution in cKPFM imaging. The use of the GP with structured kernel interpolation[29] allows reducing the model complexity and leads to more accurate predictions. We further illustrate the use of the tensor decomposition method for exploratory data analysis of the high dimensional spectroscopic imaging data that in certain cases allow separating voltage-dependent response mechanisms.

**Results and Discussion**

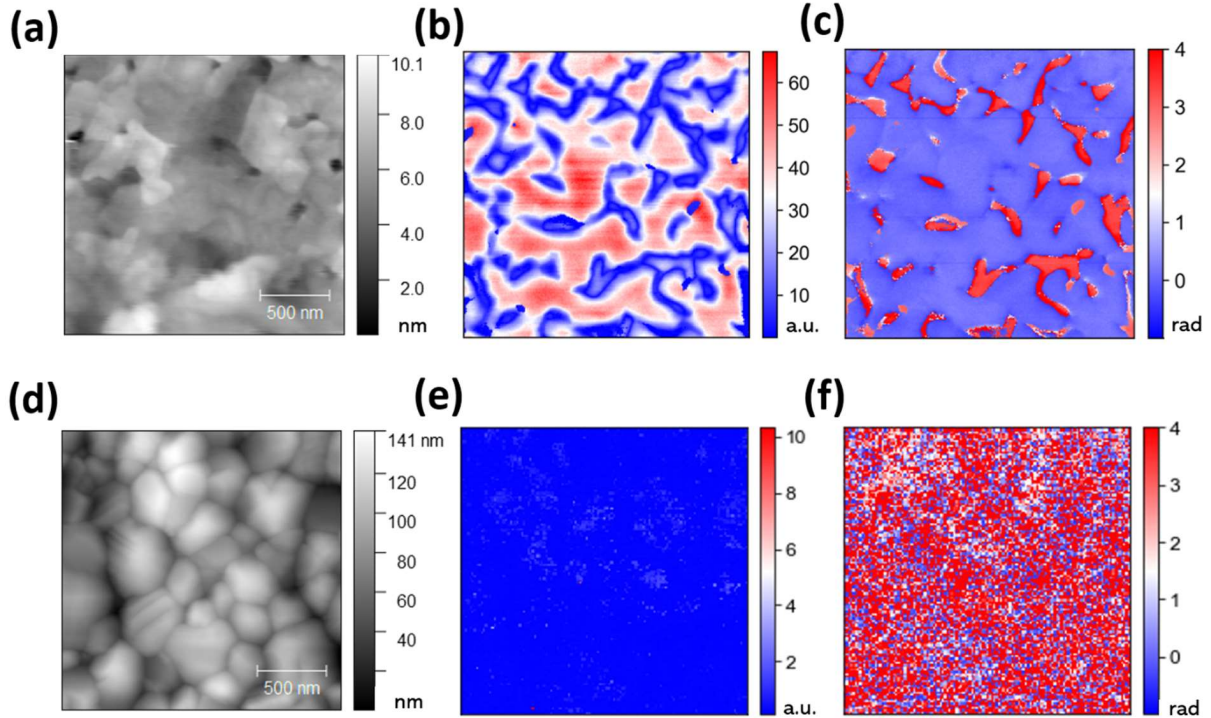

**Figure 1.** Band excitation (BE) PFM images. (a) Topographic image, (b) amplitude and (c) phase spatial maps (2 × 2 μm) of the BFO sample. (d) Topography scan, (e) amplitude, and (f) phase taken on the surface in hybrid perovskites.

To avoid topographic cross-talk, PFM measurements are performed band excitation (BE) mode for bismuth ferrite (BFO) and metal halide perovskites[30] (85FAPbI$_3$ 15MAPbBr$_3$/SnO$_2$/FTO glass) thin films in dark condition, as shown in Fig 1. The topography of the BFO surface is presented in Fig. 1(a). The corresponding amplitude and phase response are shown in Fig. 1(b) and (c) after the BE data sets are fitted with a simple harmonic oscillator (SHO) model.[22] These images show clearly ferroelectric domains as normally observed for this material. Similarly, the topography map of hybrid perovskite is shown in Fig 1(d). The corresponding amplitude and phase maps derived using the SHO model are shown in Fig 1(e) and (f). These images have significantly weaker signals compared to BFO thin film, indistinguishable from noise. This behavior can be attributed to the high leakage current and light and bias induced ionic conductivity, as well as smallness of electromechanical response compared to BFO. In fact, earlier studies using PFM techniques on MAPbI$_3$ perovskites have shown domain structures.[31-35] However, mixed compositions from compositional engineering[36] for higher power conversion efficiency (PCE, over 18%) do not have clear observation on domain structures yet using BE mode. Nevertheless, recently published other studies have shown piezo responses using normal PFM with soft cantilever[37] and DART mode[38]. This will be not discussed in detail in this paper.

To probe the coupled polarization and electrochemical behaviors in these materials, we use the contact Kelvin Probe Force Microscopy (cKPFM) approach.[28] cKPFM represents a generalization of the classical piezoresponse force spectroscopy (PFS), in which multiple hysteresis loops are measured while varying the probe voltage during the read-out process. In this manner, the polarization state set by the high-voltage pulse is measured for different tip bias values, allowing the determination of the response vs. bias. The latter can be used to determine the equilibrium surface potential corresponding to the zero-force condition. Varying the write bias in turn changes the polarization and electrochemical state of the surface. Compared to regular PFS, cKPFM allows separation of the electrostatic force and electromechanical effects in the form of zero potential loops and junction contact potential difference (jCPD, potential corresponding to the zero electrostatic force between the tip and the surface) loops, which show the read voltage at which the response is zero as a function of write voltage. While the detailed studies of the cKPFM mechanisms are still not available, it was generally found that the non-ferroelectric materials tend to yield the single band of response curves on cKPFM images corresponding to electrochemical surface charging, whereas classical ferroelectrics yield the combination of polarization hysteresis and surface charging bands, suggesting the potential for separation of phenomena associated with short range (electrochemistry, local polarization) and long range (ferroelectricity) dipoles. However, while cKPFM measurements are available for several materials classes, the systematic workflows for data analysis have still been lacking and are explored here.

The cKPFM measurements were performed on an Asylum Research AFM (MFP-3D) based on a NI (National Instrument) data acquisition card to generate probing signals and collect data using the LabView software. Fig 2(a) illustrates the cKPFM waveform in the measurements on the surface of both films using Pt-coated tips. An applied DC write voltage cycle firstly goes from zero to + 5 V, and then goes down to – 5 V, and back to 0 V sequentially for the BFO thin film. For the hybrid perovskite thin film, the write cycle goes from zero to + 2 V, and then down to – 2 V, and back to 0 V. The voltage applied during the read steps between the write pulses are from -5 V to +5 V for BFO and -2 V to +2 V for hybrid perovskites. Therefore, the cKPFM waveform generates multiple hysteresis loops with linearly increased read step voltage as shown in Fig 2(a). All measurements are based on band excitation techniques and are measured with contact resonance peaks which are fitted to a SHO model to extract data sets.

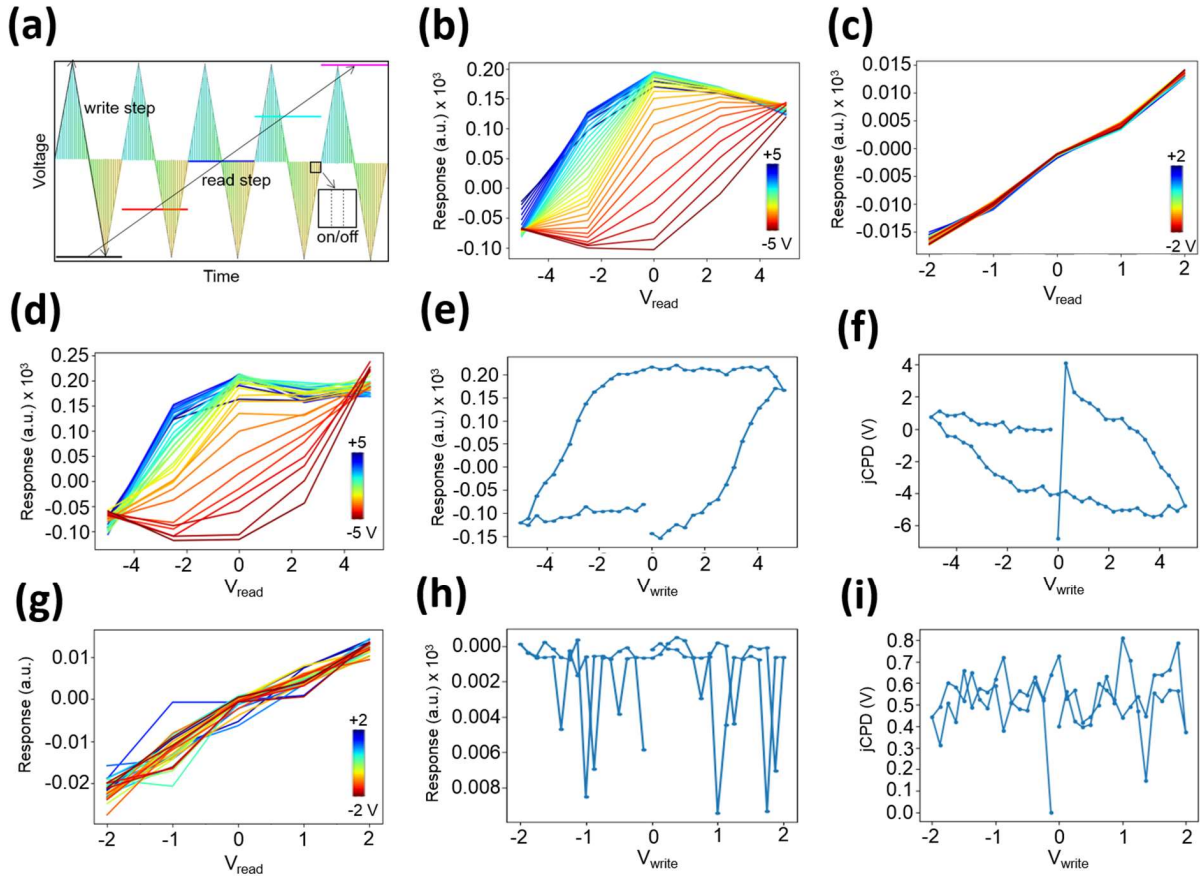

**Figure 2.** cKPFM measurements on BFO and hybrid perovskites. (a) The schematic illustration of cKPFM waveform, (b) The averaged cKPFM response as a function of read bias in BFO and (c) hybrid perovskites. The preceding write voltage for each line is indicated by the line color. (d) cKPFM response extracted from the specific point, (e) the corresponding PFM response, and (f) jCPD plot as a function write bias in BFO film. (g) cKPFM response at the specific point, (h) the corresponding PFM response, and (i) jCPD plot with write bias in hybrid perovskites.

As can be seen in Fig. 2(b) and (c), the cKPFM result on the surface in BFO film that has ferroelectric domain structures shows typically hysteretic cKPFM loop[28] with a larger band dispersion while the cKPFM response on the surface of the hybrid perovskites has a relatively linear slope with smaller band dispersion but it has trivial different behavior at positive and negative biases due to ionic motion.[39] One of the cKPFM, jCPD, and PFM response curves in BFO film show clearly nonlinear and has a typical hysteresis loop as shown in Fig 2(d-f). Here, the hysteresis loop is not fully saturated, and hence the band of ionic states over single domain region is not observed. The observed remnant offset at 0 V results from the switched dipole moments in this material which can be an indicator for ferroelectric material. However, cKPFM curve in hybrid perovskites is linear and its jCPD and PFM response are significantly weaker as well as noisy than BFO as shown in Fig 2(h-i). Note that smaller band dispersion is originated from charge carriers and ionic motion in this material. This means that hybrid perovskite can be a non-ferroelectric material but it is still debate issue which is not discussed in this study.

As noted above, one of the limitations of cKPFM measurements to date is a relatively low pixel density in space and parameter space, due to the limit on the detection bandwidth and high dimensionality of the measurement parameter space. This significantly affects the data analysis in cKPFM, since human interpretation of the observed features in the cKPFM maps and dimensionally reduced representations requires sufficiently high number of spatial pixels. Here, we explore the use of the Gaussian Process methods to interpolate the cKPFM data sets in the parameter space, enabling supper-resolution imaging.

Generally, Gaussian process is a distribution over functions on a given domain fully specified by a mean and covariance function.[40] It can be used for approximating continuous non-linear functions from a finite (sparse) number of observations. As such it can be used for recovering sparse and/or corrupted signal, as was recently demonstrated by some of the authors for the band-excitation PFM,[41] and potentially even enhancing the resolution of multi-dimensional datasets.[42,43] Here we used GP with structured kernel interpolation for both cleaning of multi-dimensional datasets and for achieving "super-resolution". We note that due to their generalization capabilities GP outperform standard nearest-neighbor or linear regression models for tasks such as image cleaning and at the same time they do not suffer from a "black box" problem of deep neural networks (the GP kernel parameters are interpretable and can be learned from the data) for tasks involving enhancement of data resolution. The major disadvantage of GPs for this type of tasks is that they are computationally expensive, having the $O(N^3)$ model complexity (*i.e.*, the computational time is proportional to the cube of the input size). One potential solution are the inducing points-based sparse GP methods that use only a subset of datapoints ($m << n$) for GP inference, which allows the model complexity to be decreased to $O(N^2)$. However, we found that in order to have a reasonable accuracy in GP predictions for multidimensional data, the number of inducing points should be at least ~5% of the overall data points, which for experimental datasets with more than 100,000 points will not fit into a GPU memory. The structured kernel interpolation framework developed by Wilson et al.[29] allows overcoming these limitations by assuming that data lies on a (partial) grid and using a combination of spectral mixture kernels and Kronecker algebra to reduce the $O(N^2)$ complexity of the sparse GP to $O(N)$ complexity. As a result, it allows a significant increase in the number of inducing points compared to regular sparse GP methods (in principle, one can use all the data points as inducing points), which leads to a greater accuracy of GP predictions. We also note that the choice of base kernel, which is "wrapped" into a structural kernel, can be guided by the physics of the problem and that priors and constraints on kernel parameters such as length scale can be obtained based on the domain knowledge about mechanisms of the processes observed. Here we used a structured kernel interpolation framework implemented in the Gpytorch library[44]. The entire paper's analysis can be retraced via the accompanying Google Colab notebook.[45]

Shown in Figure 3(a-c) is the GP processing of the cKPFM data set measured on BFO, illustrating the interpolation of the signal simultaneously in the spatial domain and the measurement parameter space. Here, we learned the GP model (kernel) hyperparameters using the exact marginal log likelihood as a "loss" function. We first applied the trained model to the original grid for smoothing/cleaning the raw 4D data (Fig. 3b). We then used the same model to predict cKPFM response for a denser grid with twice more points in all the four dimensions, which allowed us obtaining a higher resolution cKPFM dataset (Fig. 3c). The GP-processing of the cKPFM data set taken on hybrid perovskites is as shown in Fig 3(d-f). Here, small PFM responses measured dominated by noise in the original experimental data are significantly improved via GP-processing (Fig 3e, f), allowing us to see PFM responses as a function write voltage. Although Figure 3(e, f) reveals that remanent polarization is at 0 V bias, it is indicative of comparably smaller responses with less than 3 orders (see Figure 2e,h). Therefore, it shows mostly non-ferroelectric behaviour compared with a typical ferroelectric BFO sample.

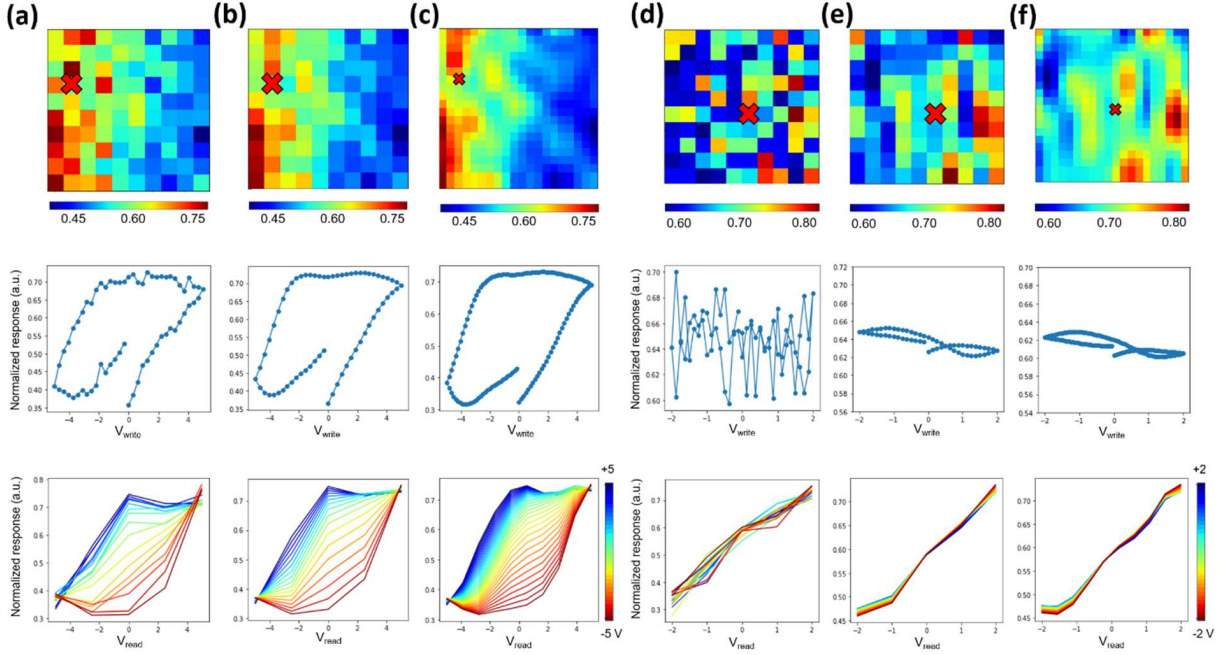

**Figure 3.** GP prediction of cKPFM data on BFO and hybrid perovskites (a) Original (raw) experimental data, (b) GP-processed data with the same resolution as the original data, and (c) GP prediction for the image with 2x resolution enhancement in all 4 dimensions in BFO. The top row shows individual slices of 4D data at fixed $V_{write}$ and $V_{read}$, while the middle and bottom rows show spectroscopic curves from the locations denoted by 'x' in the top row. The accompanying Jupyter notebook allows inspection of the full datasets. (d-f) The same processed series of data in hybrid perovskites. Notice that the response values were normalized to [0, 1] before applying the GP routines.

Both the original cKPFM data and the GP interpolation still represent the 4D data set of response as a function of position, read and write voltages, $R(x,y,V_{write}, V_{read})$, necessitating both exploratory data analysis and possible search for physically relevant representations. Point spectra and 2D representation offer an obvious way for data exploration, and in this case the GP processing significantly improves data presentation. Similarly, the classical multivariate techniques such as principal component analysis, matrix factorization, etc. can be used to separate the 4D data set into the linear combination of 2D endmembers,

$$R(x,y,V_{write},V_{read}) = \sum_{i=1}^{n} a_i(x,y) w_i(V_{write},V_{read}) \quad (1)$$

where $a_i(x,y)$ are the loading maps defining spatial variability of specific behaviors, and $w_i(V_{write}, V_{read})$ are the end members defining the behaviors. The number of the components is either chosen based on a predefined truncation criterion, or can be predefined and then optimal number can be established based on the behaviors of end members or loading maps. The loading maps can be directly visualized as 2D images. The dissimilar multivariate methods allow to impose limitations on the end members or loading maps (non-negativity, sum to one, sparsity, etc.), allowing to add the knowledge of specific physics of the system. These methods are realized in the accompanying notebook and are also explored in Ref.[46]

Here, we explore the tensor factorization (TF)[47] to derive the low dimensional representation of the cKPFM data and potentially elucidate the associated mechanisms. In TF, the 4D cKPFM data set is represented as

$$R(x, y, V_{write}, V_{read}) = \sum_{j=1}^{n} a_j(x,y) f_j(V_{write}) g_j(V_{read}), \qquad (2)$$

i.e. as a reduced representation via outer product of 1D functions.

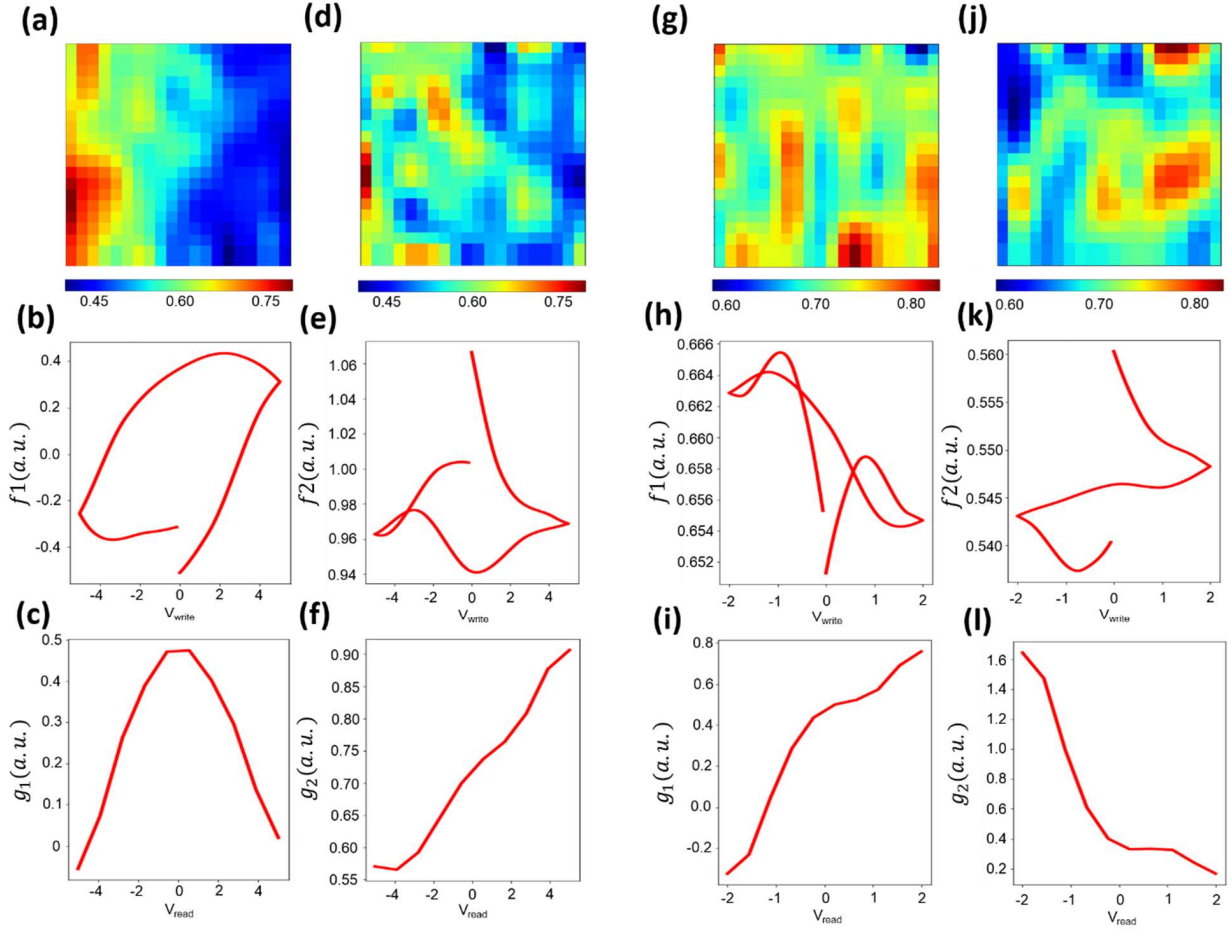

**Figure 4.** Tensor factorization of the cKPFM data. (a, d) The corresponding loading maps, (b-c, e-f) time and voltage functions for decomposition in 2 components from cKPFM data in BFO. (g, j) The corresponding loading maps, (h-i, k-l) time and voltage functions for decomposition in 2 components from cKPFM data in hybrid perovskites.

Shown in Figure 4 are the result of the TF analysis on the cKPFM data. Here, we explore the parsimonious decomposition in 2 components, albeit the order of decomposition can be varied (see the attached notebook). The TF of cKPFM data for BFO are shown in Fig. 4 (a-f). Here, the decomposition produces two pair of functional behaviors that allow straightforward physical interpretation. The first pair describes the classical ferroelectric-like hysteretic process, where $f_1(V_{write})$ follows classical hysteretic curve spanning the range of polarization values, whereas the $g_1(V_{read})$ is maximal at Vread ~ 0 and decays to the 0 close to the edges of measurement interval. This behavior is qualitatively consistent with the classical ferroelectric behavior, where read-out at high voltages "overwrites" the state of the system established by $V_{write}$. The second pair of functions shows complementary dynamics, where $f_2(V_{write})$ is almost constant (note the vertical scale in Fig. 4 e), and $g_2(V_{read})$ is a linear function of $V_{read}$. This behavior is consistent with the electrostatic like responses. Note that similar to all information theory-based methods such as non-negative matrix factorization, principal component analysis, etc. the TF decomposition does not have direct physical meaning. However, once the physics of the system comply with the assumptions made during information theory analysis, the extracted parsimonious behaviors can be analyzed as performed above.

In comparison, shown in Fig. 4 (g-l) are the TF analysis for the HOIP data. In this case, the first pair of functions $f_1(V_{write})$, $g_1(V_{read})$ depict the linear trends, likely consistent with the dominant electrostatic origin of the signal. Note that $f_1(V_{write})$ is almost cyclostationary and changes relatively week. In comparison, the second pair of function represents the non-stationary part of the signal. Furthermore, both $g_1(V_{read})$ and $g_2(V_{read})$ show strong deviations from linear responses, suggesting the presence of the complex ionic dynamics.

**Conclusion**

To summarize, here we explore the applicability of the Gaussian Process extrapolation to achieve the super resolution in the contact Kelvin Probe Force Microscopy. This approach allows effective interpolation of the data jointly in spatial domain and measurement parameter space, allowing to construct effective 1D and 2D representations for exploratory data analysis and potentially correlative data mining. We further get insight into spatial variability of response from the spatial- and voltage structure of the kernel. Finally, we show that the tensor matrix decomposition allows efficient separation of the multidimensional date set in in low-dimensional representations, and in particular case explored here their functional behaviors comport to the known physics of the imaging process, suggesting that these can be used to identify relevant physical mechanisms.

We note that the proposed approach is universal and can be applied to other multidimensional spectroscopies including FORC measurements, etc. In all cases, GP base kernel structure is expected to provide information on the length scale of the process and the correlations in the parameter space directly related to imaging mechanisms. The low dimensional representation will allow exploratory data analysis, data compression, and may allow insight into the physics of the observed phenomena. The accompanying notebook allows the reader to retrace the analysis here and apply similar workflow for their applications.

**Acknowledgements:**

Research was conducted at the Center for Nanophase Materials Sciences, which is a DOE Office of Science User Facility (MZ, RKV, LC, SJ, SVK). Part of the BE SHO data processing and experimental setup were supported by the U.S. Department of Energy, Office of Science, Basic Energy Sciences, Materials Science and Engineering Division (SMN). D.K., M.A. acknowledge support from CNMS user facility, project # CNMS2019-272. The authors would like to thank Amit Kumar (Queen's University Belfast) and Dipanjan Mazumdar (Southern Illinois University) for providing the $BiFeO_3$ sample.

This manuscript has been authored by UT-Battelle, LLC, under Contract No. DE-AC0500OR22725 with the U.S. Department of Energy. The United States Government retains and the publisher, by accepting the article for publication, acknowledges that the United States Government retains a non-exclusive, paid-up, irrevocable, world-wide license to publish or reproduce the published form of this manuscript, or allow others to do so, for the United States Government purposes. The Department of Energy will provide public access to these results of federally sponsored research in accordance with the DOE Public Access Plan (http://energy.gov/downloads/doe-public-access-plan).

**Data availability**

All data used in this manuscript are available from the authors on request.